\documentclass[%
 reprint,
 superscriptaddress,
 amsmath,amssymb,
 pre,%
 floatfix,%
]{revtex4-2}

\usepackage{balance}
\usepackage{graphicx}%
\usepackage{dcolumn}%
\usepackage{bm}%
\usepackage{hyperref}%

\begin{document}

\preprint{APS/123-QED}

\title{\texorpdfstring{Switching between Collective Network Dynamics}{Switching between Collective Network Dynamics} driven by Time-Delay Plasticity}%

\author{Stefan Ruschel}
\affiliation{School of Mathematical Sciences, University of Nottingham, University Park, Nottingham NG7 2RD, United Kingdom}

\author{Emanuil Hristov}
\affiliation{School of Mathematical Sciences, University of Nottingham, University Park, Nottingham NG7 2RD, United Kingdom}

\author{Hil G.E. Meijer}
\affiliation{Department of Applied Mathematics, Faculty of Electrical Engineering, Mathematics and Computer Science, University of Twente, Drienerlolaan 5, 7522 NB Enschede, The Netherlands}

\author{Stephen Coombes}
\affiliation{School of Mathematical Sciences, University of Nottingham, University Park, Nottingham NG7 2RD, United Kingdom}

\author{Rachel Nicks}
\affiliation{School of Mathematical Sciences, University of Nottingham, University Park, Nottingham NG7 2RD, United Kingdom}

\date{\today}%

\begin{abstract}

We show that time-delay plasticity provides a mechanism for switching between collective states in delay-coupled oscillator networks. In particular, we consider adaptive axonal delays (AADs), motivated by activity-dependent myelination in the brain, and demonstrate that AADs drive the network slowly along branches of phase-locked states until these lose transverse stability, producing explosive relaxation oscillations between collective states. We establish the mechanism for a ring network with an explicitly averageable adaptation rule, and show that a phenomenological myelination rule on human connectome data produces the same slow--fast scenario, together with frequency selection and the selection of chimera states.
\end{abstract}

\maketitle

Real-world complex networks remain difficult to understand and control as topology and time delays jointly determine dynamics \cite{atay2004delays}.
Brain activity, for example, unfolds across multiple spatial and temporal scales, posing fundamental challenges for linking neuron-level dynamics, population activity, network structure and function \cite{rabinovich2012information}. While synaptic plasticity, a central source of complexity, modifies interaction strengths in response to presynaptic activity, plasticity may also act on the timing of interactions. In particular, activity-dependent myelination %
mediated by oligodendrocytes \cite{noori2020activity,scholz2009training}
changes signal propagation speeds thereby giving rise to adaptive axonal delays (AADs), which play an important role in neural processing and computation and pose a direction of inquiry complementary to current efforts in studying evolving coupling structure \cite{berner2023adaptive,holme2012temporal}.

To study the dynamical consequences of delay plasticity, we consider an idealization of population activity to phase models \cite{ashwin2016mathematical,breakspear2010generative}, such as the Kuramoto model, %
which are suited for analyzing collective dynamical phenomena \cite{pikovsky2001synchronization}. Such phase description is particularly convenient because oscillator frequency carries the notion of spike rate and, hence, population activity. We employ a phase description that explicitly retains interaction delays \cite{novivcenko2012phase}, rather than approximation by effective phase lags 
to capture how delays affect emergent frequencies and multistability between frequencies \cite{yeung1999time}.

Considering the effects of time delays on brain dynamics \cite{wu2022time}
explicitly can potentially explain additional dynamical complexity, from pattern formation to chaos \cite{yanchuk2017spatio}, including delay-induced waves and metastable states in neural and connectome-based oscillator networks \cite{crook1997role,ko2007effects,roberts2019metastable,petkoski2019transmission,petkoski2022normalizing,koller2024human,davis2021spontaneous}. In systems with multiple time scales, time delays have been shown to generate canards \cite{krupa2016complex,campbell2009delay},
intermittent chaotic behavior \cite{ruschel2017chaotic}, 
and purely delay-induced relaxation oscillations \cite{ruschel2019delay}, and to shape slow rhythms and function in delay-coupled neural motifs \cite{qie2025synaptic,qie2026dissecting}. In systems of coupled phase oscillators, time delays have been shown to induce (explosive) synchronization \cite{scholl2009time,peron2012explosive}, promote chimera states \cite{omel2010chimera,sethia2008clustered}, and, when adaptive, enable mode selection 
and adiabatic switching between attractors \cite{klinshov2024adaptive,karimian2019effects,selivanov2012adaptive}. These attractor-switching dynamics are analogous to network attractors in heteroclinic and excitable networks previously proposed as models for transitions between brain states \cite{ashwin2024network}.

AADs by means of adaptive conduction speeds have been modelled before. Refs.~\cite{pajevic2014role,noori2020activity} adapt link-wise velocities so as to reduce the phase mismatch between pre- and post-synaptic populations, Refs.~\cite{talidou2022homeostatic,lefebvre2025myelin} consider rate-dependent velocity plasticity in spiking and rate networks, and Ref.~\cite{jolly2026} drives each link by the presynaptic activity accumulated over its transit time. In all cases speeds settle and the network becomes more coherent.

In this Letter, we consider axonal speed adaptation by a thresholded, saturating function of the presynaptic frequency alone, which renders the adaptation independent of the postsynaptic target and admits an explicit slow flow on the manifold of phase-locked states. We establish that the resulting AADs enable the selection of, and recurrent transitions between, collective network dynamics, including synchrony, phase-locked states, clusters, and chimera states. The transitions arise from a slow--fast mechanism. Adaptation drives the network along a branch of phase-locked states until the branch loses transverse stability, the network then reorganizes explosively on the fast time scale, and the resulting recurrent cycle, rather than convergence to a single state, constitutes a network attractor. We arrived at this result by studying phase oscillators with white-matter plasticity rules on human connectome data, where recurrent transitions between collective states appeared alongside frequency and chimera selection. To understand how these transitions come about we then turned to a regular ring network with an adaptation rule that can be averaged explicitly, choosing the interaction function for which the organisation of phase-locked states is rich, and analysed the mechanism there.

Specifically, we consider a system of delay-coupled phase oscillators $\theta_i(t) \in [0, 2 \pi),$ $1\leq i \leq N,$
\begin{equation}
\theta^\prime_i(t) = \omega + \sigma \sum_{j=1}^N {w}_{ij} h(\theta_{j}(t-\tau_{ij}(t)) - \theta_i(t)), 
\label{eq:theta}
\end{equation} 
where $'={\rm d}/{\rm d}t$ and $N<\infty$.
The parameters $\omega\in\mathbb{R}$ and $\sigma\ll1$ denote the (for illustration homogeneous) natural frequency of oscillator $i$, and overall coupling strength of the network, respectively. The coupling is mediated by coupling weights $w_{ij},$ coupling delays $\tau_{ij},$ and the $2\pi$-periodic interaction function $h$. Each oscillator represents the phase of a neural population (a phase-reduced neural mass \cite{jolly2026}), so that $\theta_i'$ measures the population rhythm rather than the firing rate of an individual neuron.
We consider prototypical examples $h(\theta) = \sin(\theta+a),$ recovering the Kuramoto model for $a\!=\!0$, and bi-harmonic $h(\theta) = -\sin(\theta+a) + r\sin(2 \theta)$ supporting cluster states \cite{ashwin2016mathematical}. The lag $a$ has been used to account for synaptic and dendritic filtering \cite{sakaguchi1986soluble}. It enters the locking conditions below only through the effective lag $a-\Omega\tau_{ij}$ and therefore shifts the branches of phase-locked states without altering the mechanism we describe \cite{supp}.

The delays $\tau_{ij}$ are determined by (physical) distances $d_{ij}$ and axonal signal velocities ${\rm v}_{ij}= v_j,$ such that
\begin{equation}\label{eq:tau}
   \tau_{ij} = d_{ij} v_{j}^{-1},\quad v^\prime_j= \varepsilon g(\theta'_j,v_j),
\end{equation}
depending on  the frequency (activity) $\theta^\prime_j$ of the projecting node, and a suitable non-linear gain function $g$; thus, rendering \eqref{eq:theta}--\eqref{eq:tau} a state-dependent delay differential equation \cite{hartung2006functional} and changing the dynamics away from equilibrium \cite{humphries2022nonlinear}.
The speeds considered here are node-wise, which does not hold in general, i.e. for rules that depend on postsynaptic activity \cite{noori2020activity,talidou2022homeostatic} or on the fibre length itself \cite{jolly2026}.

\begin{figure*}[!]  \includegraphics[width=1.\linewidth]{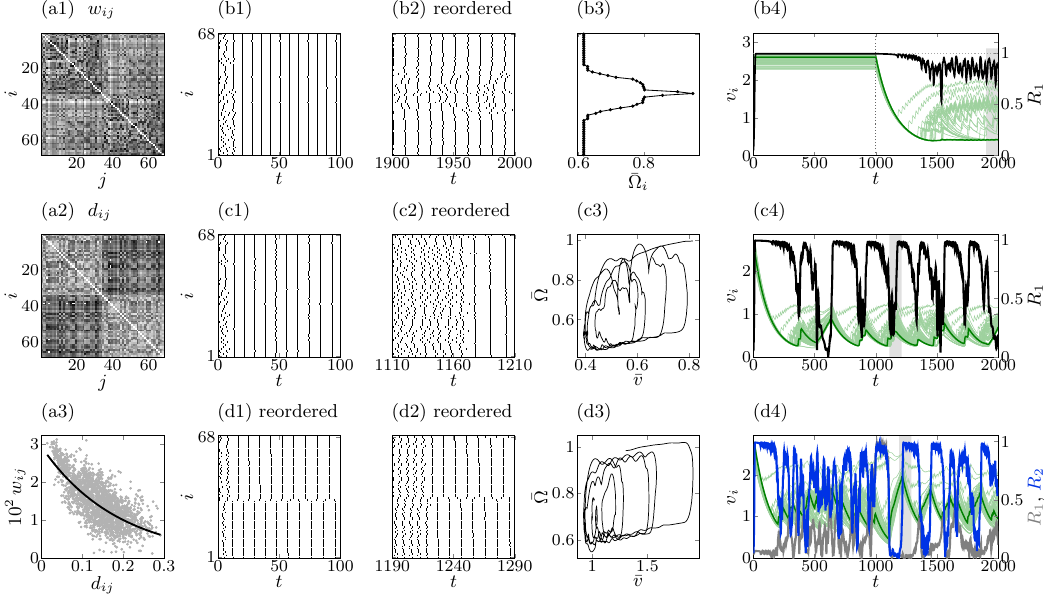}
    \caption{Selection of a chimera state (b), of switching between synchrony and incoherence (c) and of switching between cluster states (d) on the 68-node regional brain connectome (a). (a1),(a2) Coupling weights $w_{ij}$ (average row sum $1$) and distances $d_{ij}$ (mean $1$), grey scale from white ($0$) to black. (a3) Weights against raw distances with exponential fit. (b1),(c1),(d1) Raster plots for $t\in[0,100]$ from random initial conditions, with a tick whenever the phase $\theta_i$ of oscillator $i$ crosses a multiple of $2\pi$, for $h(x)=\sin x$ (b),(c) and $h(x)=c_2[\sin(x-\pi/2)+\sin 2x]$, $c_2=0.568$ (d), see Fig.~\ref{fig:4}(a). (b2),(c2),(d2) Raster plots over the $100$ time units shaded grey in (b4),(c4),(d4). Panels marked ``reordered'' list the oscillators by mean frequency with the fastest in the middle (b2), by descending $\theta_i(100)$ mod $2\pi$ (c2),(d1), or by descending phase difference to the mean phase at $t=1240$, the middle of the window (d2), and otherwise by index. (b3) Mean frequencies $\bar\Omega_i$ over $t\in[1500,2000]$ in the order of (b2). (c3),(d3) Relaxation oscillation projected onto mean speed $\bar v$ and collective frequency $\bar\Omega$ for $t\ge1150$. (b4),(c4),(d4) Speeds $v_i$ (light green, with $v_1$ dark green, left axis) and order parameters $R_1=|\sum_j {\rm e}^{{\rm i}\theta_j}|/N$ (black, grey in (d4)) and $R_2=|\sum_j {\rm e}^{2{\rm i}\theta_j}|/N$ (blue, only in (d4)), right axis. In (b) the speeds are kept constant until $t=10^3$ (dotted line). Parameters: $\omega=1$,  $\varepsilon=0.01$, $v_{\min}=0.1$, $v_{\max}=20$, $\beta=5$, (b) $\sigma=0.6$, $\nu=1.44$, (c) $\sigma=0.8$, $\nu=1.56$, (d) $\sigma=1.0$, $\nu=1.4$, marked by squares in Fig.~\ref{fig:4}(b),(g), with weights rescaled to an average row sum of $1$, so that $\sigma$ measures the total input per node, and distances to mean $1$. Initial data: $\theta_i(t)\sim\mathcal{U}(0,2\pi)$, and $v_i(t)\sim\mathcal{N}(2.5,0.1)$ restricted to $[v_{\min},v_{\max}]$.}
    \label{fig:2}
\end{figure*}

In the human brain, the myelination time scale ($1/\varepsilon$) is long, from days ($\sim 10^3$s) to months ($\sim 10^7$s) \cite{scholz2009training}, compared to the spiking time scale of isolated neurons ($\sim 10^0$s). Consequently, we can assume $\varepsilon \ll \sigma \ll 1,$ and thus, axonal speeds $v_j$ to be well approximated (up to order $\varepsilon$ on time scale $1/\varepsilon$) by the averaged velocities $V_j,$ satisfying
$$ V^\prime_j = \varepsilon\langle g(\theta'_j,V_j) \rangle = \varepsilon\lim_{T\to\infty} \frac{1}{T} \! \int_0^T  \!\! g(\theta^\prime_j(s),V_j)\,{\rm d}s.$$
Specifically, Eq.~\eqref{eq:tau} induces the slow flow  
\begin{equation}\label{eq:slowflow}
   V_j'
   =
   \varepsilon\,
   g\!\left(\Omega(V),V_j\right),
\end{equation}
on families of
phase-locked solutions $\theta_1(t)=\Omega(V)t,$
$\theta_i(t)=\Omega(V)t+\phi_i,$ $i=2,\ldots,N,$ where the collective frequency \(\Omega(V)\) and phase offsets
\(\phi=(\phi_1,\ldots,\phi_N)\), $\phi_1=0$ are implicitly determined by the locking conditions
\begin{equation}\label{eq:locking}
\Omega(V)
=
\omega
+
\sigma\sum_{j=1}^{N} w_{ij}\,
h\!\left(
\phi_j-\phi_i-d_{ij}V_j^{-1}\Omega(V)
\right),  
\end{equation}
 We therefore expect to observe attraction of the fast flow onto transversally stable branches of the locking manifold \eqref{eq:locking} followed by slow flow on that manifold as long as it remains transversally stable. Equilibria $V^\ast$ of the slow flow \eqref{eq:slowflow} on the locking manifold are determined by the algebraic conditions
$ g\!\left(\Omega(V^\ast),V_j^\ast\right)=0,$ $1\leq j \leq N.$ The scenario does not hinge on a strict separation of time scales, and its dependence on $\varepsilon$ is quantified in \cite{supp}.

\begin{figure*}[!tbp]
    \centering    \includegraphics[width=\linewidth]{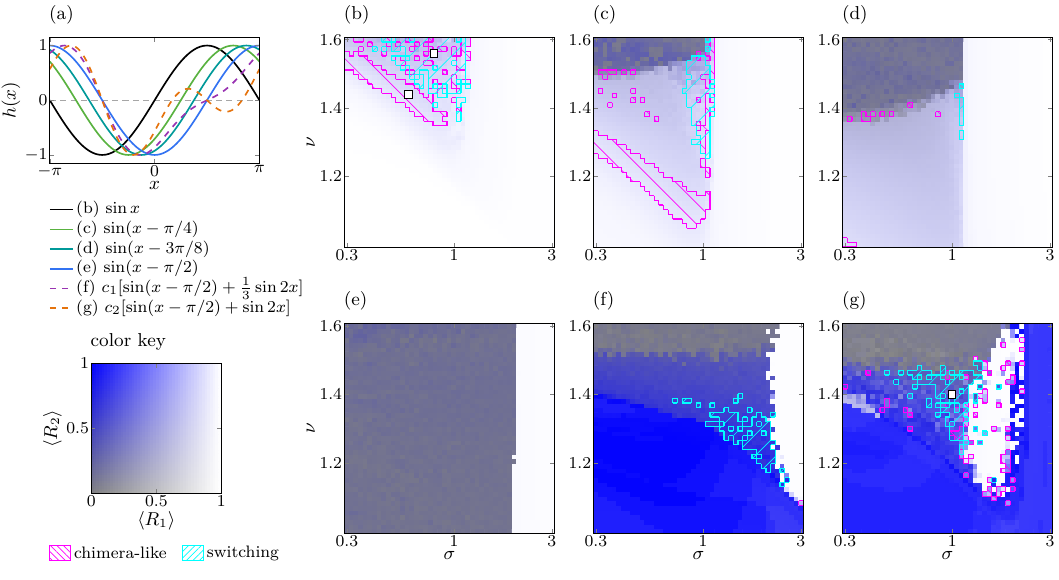}%
    \caption{Chimera states and switching across interaction functions. (a) The interaction functions considered. (b)--(g) Long-time behavior of \eqref{eq:theta}--\eqref{eq:g1} on the connectome as a function of coupling strength $\sigma$ and threshold $\nu$ for (b) $h(x)=\sin x$, (c) $h(x)=\sin(x-\pi/4)$, (d) $h(x)=\sin(x-3\pi/8)$, (e) $h(x)=\sin(x-\pi/2)$, the interaction function of the ring network of Fig.~\ref{fig:3}, (f) $h(x)=c_1[\sin(x-\pi/2)+\tfrac13\sin 2x]$ and (g) $h(x)=c_2[\sin(x-\pi/2)+\sin 2x]$, with $c_1=0.861$ and $c_2=0.568$ chosen such that $\max h=1$ ($45\times45$ grid, the same random initial condition in every cell, $t\le 1.5\times10^3$). The color shows the time averages over the second half of the runs of the Kuramoto order parameter $\langle R_1\rangle$ and of the cluster order parameter $\langle R_2\rangle$, encoded as the pair $(\langle R_1\rangle,\langle R_2\rangle)$ by the two-dimensional key below (a) (grey incoherent, white synchronous, blue two-cluster states). Cyan hatching marks recurrent switching, i.e.\ $R_1$ or $R_2$ rises above $0.9$ and collapses below $0.1$ at least twice during the second half of the run. Magenta hatching marks chimera-like states, i.e.\ a frequency-locked and a drifting group coexist (between $10\%$ and $90\%$ of the nodes share the common mean frequency, tolerance $10^{-3}$, and $0.2<\langle R_1\rangle<0.9$). Squares mark the parameters of Fig.~\ref{fig:2}(b) and (c), the lower and upper square in (b), and of Fig.~\ref{fig:2}(d), in (g). Other parameters and normalization as in Fig.~\ref{fig:2}.}
    \label{fig:4}
\end{figure*}

\paragraph*{Network attractor selection by AADs on brain connectome topology.}

We consider \eqref{eq:theta}--\eqref{eq:tau} on a network topology obtained from coarse-grained structural brain connectome data \cite{van2013wu,forrester2024whole} (68 Desikan--Killiany brain regions) with AAD rule
\begin{equation}\label{eq:g1}
\begin{aligned}
    g(\theta^\prime_j,v_j)=&- v_j + v_{\min}+  (v_{\max}\!  - \!v_{\min})\\
    &\times f\left(\int_{-\infty}^0\!\!k(t+s)\theta^\prime_j(s)\, {\rm d}s -\nu\right),
\end{aligned}
\end{equation}
where $\beta$$,v_{\min},v_{\max},\nu\geq 0,$ and $f(x)=1/(1+{\rm e}^{-\beta x}),$ $\beta>0$ (scaled sigmoid).  Equation \eqref{eq:g1} is a phenomenological rule for activity-dependent myelination. Its biologically supported ingredients are that myelination is slow, activity dependent, saturating and bounded \cite{noori2020activity,scholz2009training,talidou2022homeostatic}. The choice of the filtered presynaptic frequency $\int_{-\infty}^0\!\!k(t+s)\theta^\prime_j(s)\, {\rm d}s$ as the drive, the soft threshold $\nu$, the sigmoidal form and the values of $v_{\min},v_{\max}$ are modelling conveniences. A direct consequence, open to test, is that activity-dependent speed-up reduces the dispersion of conduction speeds across projecting regions and pulls the collective frequency towards $\nu$. Figure \ref{fig:2}(a) illustrates the connectome data showing (in gray scale) the normalized pairwise coupling weights (a1), normalised distance (a2), and exponential least squares fit (exponent $-5.34$) of coupling weights as a function of distance (a3). The coupling weights in Fig.~\ref{fig:2}(a) have been rescaled so that the average row sum is $1$ in arbitrary units, i.e.\ $\sigma$ measures the total input per node. Overall, the system is fully coupled with distance-dependent coupling with some degree of heterogeneity.

Figure~\ref{fig:2}(b) shows a numerical simulation of system \eqref{eq:theta} with interaction function $h(x)=\sin(x),$ memory kernel $k=10\cdot\mathbf1_{[-0.1,0]}$ (the mean frequency over the preceding $0.1$ time units), and (constant) random initial conditions for the phases and nearly identical initial speeds. The delays are kept constant from $t=0$ to $t=10^3$ before switching to the AAD rule \eqref{eq:tau},\eqref{eq:g1}. Panels (b1),(b4) show that oscillators quickly synchronize when the  delays are kept constant and remain synchronized until delays are allowed to vary. In response to switching on the adaptivity, the speeds decrease and spread out, since the collective frequency lies below the threshold $\nu$. In this case, AADs select a chimera state; cf. Fig.~\ref{fig:2}(b2) (oscillators reordered by average frequency such that the fast, incoherent part is centered). Panel (b3) illustrates the partial frequency locking of the ensemble through the average frequency of oscillators (reordered).  The chimera state persists for the full duration of our simulations ($t\le 10^6$, i.e.\ $10^4$ adaptation times $1/\varepsilon$). The network neither synchronizes nor desynchronizes completely ($0.84<\langle R_1\rangle<0.86$ in every window of $10^4$ time units), while the membership of the frequency-locked group slowly changes \cite{supp}. Chimeras of identical oscillators with a single effective phase lag \cite{abrams2004chimera,omel2010chimera,laing2009chimera} are typically chaotic transients at finite $N$ \cite{wolfrum2011chimera}. Here, although the speeds homogenize, the delays $\tau_{ij}=d_{ij}/v_j$ remain heterogeneous and the coupling topology is not homogeneous, so the system does not reduce to a single phase lag.

Raising the threshold to $\nu=1.56$ (with $\sigma=0.8$ and the same interaction function $h(x)=\sin x$) replaces the chimera by recurrent switching between synchrony and incoherence, now with adaptation from $t=0$, as shown in Fig.~\ref{fig:2}(c). Panel (c1) shows that the oscillators synchronize quickly from random initial phases. Along the synchronous state the speeds decrease and the collective frequency falls to about $0.45$. Synchrony is then lost, and in the ensuing incoherent transient the frequency rises towards $\nu$, the speeds grow and the network resynchronizes, with a period of about $220$ time units (panels (c2) and (c4)). In the plane of mean speed and collective frequency, panel (c3), a slow drift at low frequency alternates with a fast excursion to high frequency.

Figure~\ref{fig:2}(d) shows a qualitatively different behavior, namely relaxation oscillations between different phase-locked states with normalized interaction function $h(x)=c_2[\sin(x-\pi/2)+\sin 2x]$, $c_2=0.568$, whose second harmonic promotes multi-stability of phase locked states and cluster states \cite{ashwin2016mathematical}. More precisely, we observe a slow decrease of the speeds along a two-cluster state, with $R_2$ close to one and the collective frequency below $\nu$ (panels (d2) and (d4)), its breakdown into a transient incoherent state in which the collective frequency rises towards $\nu$ and the speeds increase, and the subsequent reorganization into a different two-cluster state, repeated irregularly every $100$--$250$ time units. Recurrent transitions between a few discrete patterns of functional connectivity are observed in resting-state fMRI and MEG \cite{allen2014tracking,vidaurre2018spontaneous}. In phase-based analyses each pattern divides the brain regions into two groups by their relative phase \cite{cabral2017cognitive}, much like the two-cluster states found here. Comparable transitions arise in whole-brain network models \cite{hansen2015functional,cabral2014exploring}, where conduction delays shape the patterns of partial synchronization \cite{cabral2014exploring,petkoski2019transmission}. The present model is far from a realistic large-scale neural model, and the analogy is qualitative.

The transitions in Fig.~\ref{fig:2}(c) and (d) follow the slow--fast scenario analyzed below for the ring. Along each phase-locked state the speeds drift slowly and the collective frequency decreases until the state loses transverse stability and the network reorganizes on the fast time scale. Figure~\ref{fig:2}(c3),(d3) make this explicit by projecting the trajectories onto the plane of mean speed and mean frequency, in which a slow drift at low frequency alternates with a fast excursion to high frequency. The selected states are not confined to specific parameter values. Figure~\ref{fig:4} summarizes the long-time behavior over a range of coupling strengths $\sigma$ and thresholds $\nu$ and over random initial conditions for six interaction functions. Partially frequency-locked (chimera-like) states in panels (b) and (c) and recurrent switching in panels (b), (c), (f) and (g) each occupy extended regions, which shrink as the phase lag grows (d), whereas with the phase lag $\pi/2$ of the ring alone (e) the heterogeneous delays leave the connectome incoherent unless the coupling is strong enough to synchronize it. The dependence on $\beta$, $\varepsilon$ and the spread of initial speeds is reported in \cite{supp}. The same scenario is obtained with the polynomial rule \eqref{eq:g2} on the connectome (for thresholds near the upper end of the attained collective frequencies and slow adaptation) \cite{supp}.

\begin{figure*}
    \centering    \includegraphics[width=\linewidth]{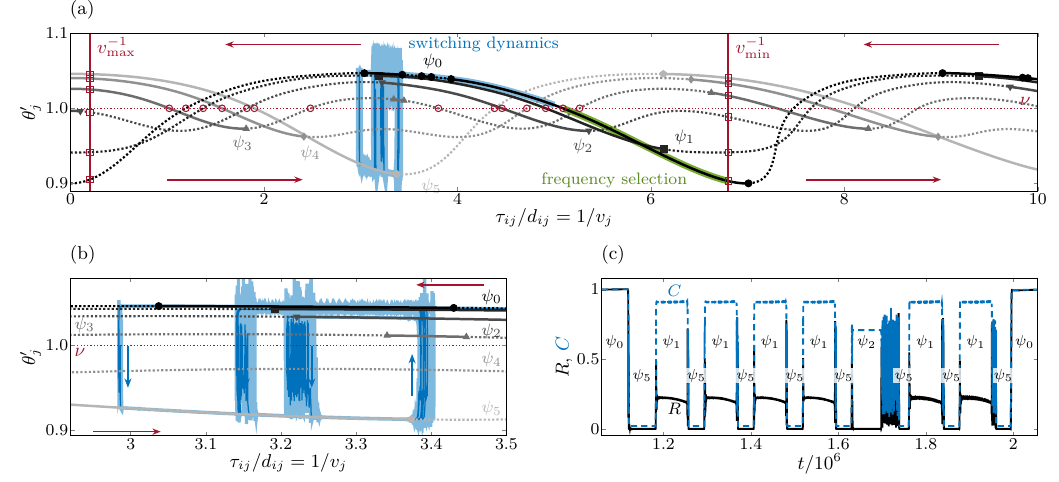}%
    \caption{Frequency selection (green) and switching dynamics (blue) in spatially embedded, non-local ring graph with $N=11$ nodes, interaction function $h(x)=\sin(x-\pi/2),$ and gain function \eqref{eq:g2}. (a) greys: bifurcation diagram ($\varepsilon=0$) of stable (solid) and unstable (dotted) $m$-twisted states with phase difference $\psi_m=m2\pi/11,$ (labelled in the panels) with respect to (constant) delays AAD $\tau_n=1/ v_n$ (markers correspond to bifurcation points); reds: threshold values $\nu$ (red dotted, open circles), $v_{\min}^{-1}, v_{\max}^{-1}$ (red solid, open squares) with markers corresponding to equilibria of the slow flow; blue/green: projection of time series onto $(\theta_n^\prime,\tau_n)$-plane (light blue: node $1$, dark blue: network mean); (b) Enlargement of (a), arrows indicate direction of the slow (red) and fast (blue) flow; (c) evolution of Kuramoto order parameter $R$ (black solid), and mean nearest-neighbor coherence $C=\frac{1}{N}\sum_{i=1}^{N}F_{i,i+1}$, $F_{N,N+1}\equiv F_{N,1}$ (blue dashed) along blue trajectory for one cycle from the departure from $\psi_0$ until the return to it, with the twisted state held in each segment indicated. Parameters: $\omega=1$, $\sigma=0.1$, $\varepsilon=10^{-5}$, $\nu=1$, $v_{\min}=1/6.8$, $v_{\max}=5$. Initial data $\theta_j(t)=\Omega t$ and $v_j(t)=c$  with (blue) $\Omega=1.0004$, $c=5$, (green) $\Omega=0.9996$, $c=5.1$.}
    \label{fig:3}
\end{figure*}

\paragraph*{Fully-coupled ring network with exponentially decaying weights.}

The connectome topology is little amenable to studying the transitions of Figs.~\ref{fig:2}(c),(d). To isolate the mechanism we therefore consider a fully coupled ring network with distances $d_{ij}=\min \{|i-j|,N-|i-j|\},$ and weights $w_{ij}=\exp(- {\rm dist}(i,j))$ for $i\neq j,$ and $w_{ij}=0$ (no self connections, and only depending on the difference of indices), cf.\ Ref.~\cite{jolly2026}. We choose $N=11$ moderately large and prime to avoid many symmetries and consider the idealization (which can be averaged explicitly)
\begin{equation}
\label{eq:g2}
    g(\theta^\prime_j,v_j)=(\theta^\prime_j-\nu)(v_j-v_{\min})(v_{\max}-v_j)
\end{equation}
with (common) parameters $v_{\min},v_{\max}>0$ and $\nu\in\mathbb{R}.$ 
The plasticity rule \eqref{eq:tau},\eqref{eq:g2} renders the speed dynamics sensitive to frequency pulling and collective phase locking by $v_j'(t) \geq 0$ ($\leq 0$) when $\langle \theta^\prime_j(t)\rangle \geq \nu$ ($\leq \nu$) and $v_{\min}\leq v_j(t) \leq v_{\max}$ with explicit (homogeneous) equilibria at the intersection of each branch with the lines $\Omega(v)=\nu,$ and $v_j=v_{\min},$ and $v_j=v_{\max}.$

Linearizing about an $m$-twisted state $\theta_i=\Omega t+i\psi_m$ of the fast subsystem with frozen delays ($\varepsilon=0$) with uniform speeds $v_j=v_1$ and diagonalizing by the discrete Fourier transform \cite{jolly2026} yields, for each mode $q_n=2\pi n/N$, $n=0,1,\ldots,N-1$, the scalar delay equation $\dot\eta=-\sum_j c_j\,\eta(t)+\sum_j c_j{\rm e}^{{\rm i}q_n j}\eta(t-\tau_{0j}),$ $j=1,\ldots,N$ with $c_j=\sigma w_j h'(j\psi_m-\Omega\tau_{0j})$, and $\tau_{0j}=d_{0j}/v_1$. By Corollary~7 of Ref.~\cite{yanchuk2022absolute} every such mode is stable for all delays, and the twisted state is absolutely transversally stable, if $c_j\ge0$ for all $j$ (up to a nondegeneracy condition \cite{supp}). Differentiating the locking condition \eqref{eq:locking} gives
\begin{equation}\label{eq:slope}
\frac{{\rm d}\Omega}{{\rm d}(1/v_1)}
=-\frac{\Omega\sum_j d_{0j} c_j}{1+v_1^{-1}\sum_j d_{0j} c_j}<0,
\end{equation}
so that absolutely stable families of m-twisted states have negative slope when plotted with respect to $1/v_1$. The converse does not hold, cf.\ Ref.~\cite{jolly2026}. As a result, any equilibrium of the slow flow with $\Omega(v)=\nu$ on such a branch is longitudinally unstable, and only the boundary states $v_j=v_{\min}$ and $v_j=v_{\max}$ can be selected.

This mechanism is illustrated in Fig.~\ref{fig:3}(a), where a static bifurcation diagram of phase-locked states (grey) is overlaid with simulations initialized on the synchronous branch $\psi_0$ at different initial speeds, $v_n(t)=5$ (blue) and $v_n(t)=5.1$ (green). The corresponding frequencies, $\Omega(1/5)\approx1.0004$ and $\Omega(1/5.1)\approx0.9996$, were chosen to lie just above and below the threshold $\nu=1$, respectively. For $\Omega(1/5.1)<1$ (green), the trajectory drifts slowly to the right until frequency selection arrests the dynamics on the synchronous branch at $V=v_{\min}$. Conversely, for $\Omega(1/5)>1$ (blue), the trajectory drifts leftward along $\psi_0$ until this branch loses transverse stability at the leftmost filled black circle. The ensuing dynamics consist of slow phase-locked segments, indicated by red arrows, separated by fast transition layers, indicated by blue arrows, and remain confined to a small region of parameter space; see the enlargement in Fig.~\ref{fig:3}(b). Figure~\ref{fig:3}(c) shows the 16 explosive transitions of this relaxation oscillation after departure from $\psi_0$ until the return to it, during which the trajectory visits the splay branch $\psi_1$ and the twisted branches $\psi_2$ and $\psi_5$ before returning to $\psi_0$. The resulting attractor of the full system is the entire relaxation cycle. The phase-locked states visited along it are invariant sets of the fast ($\varepsilon=0$) subsystem that become metastable once drift is switched on. In the terminology of Ref.~\cite{ashwin2024network} this is a network attractor whose transitions between collective states are driven deterministically by the slow variable rather than by noise, a realization of winnerless competition induced by delay plasticity, akin to the slow switching through delayed coupling in Ref.~\cite{kori2001slow}. The cycle persists when the time-scale separation is weakened up to $\varepsilon=10^{-3}=10^{-2}\sigma$. For larger $\varepsilon$ the trajectory no longer tracks the locking branches \cite{supp}.

\paragraph*{Summary and outlook.}

Time-delay plasticity provides a mechanism for shaping collective behavior in adaptive networks that is distinct from, and complementary to, synaptic plasticity. Here we have shown that AADs can generate network attractors in the form of explosive relaxation oscillations, revealing an intrinsically nonlinear route by which adaptive delays organize collective activity. More broadly, adaptive state-dependent delays can create, select, and stabilize collective rhythms, and the same scenario, slow drift along phase-locked branches interrupted by fast switching, has recently been observed with a spiking white-matter plasticity rule \cite{coombes2026adaptive}, indicating that it is not an artefact of the phase description. A systematic account of phase-locked states under state-dependent delays is given in Ref.~\cite{jolly2026}. Whether experimentally constrained rules for white-matter plasticity fall into the same class remains to be seen.

Several directions follow naturally. First, extending the present neural-mass and phase-reduced descriptions to higher-order phase reductions \cite{nicks2024phase} could capture regimes beyond weak coupling. Second, spiking-neuron models with biologically realistic myelination dynamics, resource constraints, spatial coupling kernels, and heterogeneity in initial conditions, intrinsic frequencies, or adaptation rates should clarify how delay plasticity controls collective states \cite{coombes2023neurodynamics}. In the strong-coupling regime, delay-coupled excitable or spiking units can exhibit collective spiking patterns ranging from traveling waves to bursts \cite{klinshov2015multistable}, motivating extensions of master-stability theory for traveling waves \cite{ruschel2025master} to delay-coupled systems. Finally, extending mean-field reduction techniques \cite{bick2020understanding} to adaptive delays offers a route toward low-dimensional descriptions of collective states.

The rigorous analysis of delay-induced relaxation oscillations induced by time-delay plasticity requires 
further development of infinite-dimensional Fenichel theory \cite{bates1998existence}, including blow up techniques. Relevant nonlinear mechanisms include slow passage through Hopf bifurcations, canards, and mixed-mode oscillations \cite{baer1989slow,avitabile2020local,desroches2012mixed}. 

\paragraph*{Acknowledgments.}
Research funded by the Leverhulme Trust Research Project Grant no. RPG-2025-052. Data were provided in part by the Human Connectome Project, WU-Minn Consortium (Principal Investigators: David Van Essen and Kamil Ugurbil; 1U54MH091657) funded by the 16 NIH Institutes and Centers that support the NIH Blueprint for Neuroscience Research; and by the McDonnell Center for Systems Neuroscience at Washington University.
\paragraph*{Code and Data Availability.}
See Supplemental Material \cite{supp} for the derivation of the stability criterion, the comparison of node-wise and link-wise speeds, and parameter sweeps. All code and data is freely available at \cite{github}. Numerical simulations are performed with 
\textsc{DifferentialEquations.jl}  %
in \textsc{Julia} and the stability of phase-locked states has been computed using \textsc{DDE-Biftool} \cite{sieber2014dde} 
 for \textsc{Matlab}.  

\balance

\onecolumngrid
\clearpage
\begin{center}
{\large\bfseries Supplemental Material:\\[2pt] Switching between Collective Network Dynamics driven by Time-Delay Plasticity\par}
\vspace{6pt}
Stefan Ruschel, Emanuil Hristov, Hil G.E. Meijer, Stephen Coombes, and Rachel Nicks\par
\vspace{12pt}
\end{center}
\twocolumngrid
\setcounter{section}{0}
\setcounter{equation}{0}
\setcounter{figure}{0}
\setcounter{table}{0}
\renewcommand{\thesection}{S\arabic{section}}
\renewcommand{\theequation}{S\arabic{equation}}
\renewcommand{\thefigure}{S\arabic{figure}}
\renewcommand{\thetable}{S\arabic{table}}
\renewcommand{\theHsection}{S\arabic{section}}
\renewcommand{\theHequation}{S\arabic{equation}}
\renewcommand{\theHfigure}{S\arabic{figure}}
\renewcommand{\theHtable}{S\arabic{table}}
% the reference list of the Supplemental Material records its width under its own label (revtex uses LastBibItem)
\makeatletter
\def\NAT@bibsetnum#1{\setlength{\topsep}{\z@}\NATx@bibsetnum{\ref{LastBibItemSM}}}%
\renewenvironment{thebibliography}[1]{%
 \NAT@thebibliography{#1}\let\@TBN@opr\present@bibnote\@FMN@list
}{%
 \auto@bib@innerbib\edef\@currentlabel{\arabic{NAT@ctr}}\label{LastBibItemSM}%
 \endNAT@thebibliography\aftergroup\auto@bib@empty
}
\makeatother

This Supplemental Material contains (S1) the derivation of the transverse stability criterion and of the slope relation used in the Letter, (S2) the reduction from link-wise to node-wise speeds, (S3) the dependence of the ring dynamics on the time-scale separation $\varepsilon$ and on the phase lag, and (S4) the connectome data and the numerical sweeps behind Figs.~1 and 2 of the Letter. Equation numbers without prefix refer to the Letter.

\section{Transverse stability of twisted states}

We consider phase-locked (twisted) states of the delay-coupled ring network of the Letter with constant speed $v$ and delays $\tau_j=d_j/v$, $d_j=\min\{j,N-j\}$, $w_{ij}=w_{d_j}$ (writing $w_j$ for short). The $m$-twisted state $\theta_i(t)=\Omega t+i\psi_m$, $\psi_m=2\pi m/N$, has collective frequency determined by
\begin{equation}
\Omega=\omega+\sigma\sum_{j=1}^{N-1} w_j\,h\!\left(j\psi_m-\tau_j\Omega\right).
\label{eq:freq_lock}
\end{equation}
The construction of such states, their linear stability through the characteristic function, and the diagonalization by the discrete Fourier transform for circulant coupling are given in Ref.~\cite{S-jolly2026}, Secs.~III--IV, which also contains the parametric construction used for the static bifurcation diagram in Fig.~3(a) of the Letter (their Appendix~B.2). Here we only record the steps that lead to the delay-independent criterion.

Linearizing about the twisted state and writing perturbations as $\delta\theta_i(t)=\eta_n(t)e^{{\rm i} q_n i}$, $q_n=2\pi n/N$, $n=0,\dots,N-1$, each Fourier mode evolves independently according to the scalar delay equation
\begin{equation}
\dot{\eta}_n(t)=a_{n,0}\,\eta_n(t)+\sum_{j=1}^{N-1} a_{n,j}\,\eta_n(t-\tau_j),
\label{eq:scalar_mode}
\end{equation}
with
\begin{equation}
\begin{aligned}
a_{n,0}&=-\sum_{j} c_j, \qquad a_{n,j}=c_j\,{\rm e}^{{\rm i} q_n j},\\
c_j&:=\sigma w_j\,h'\!\left(j\psi_m-\Omega\tau_j\right).
\end{aligned}
\label{eq:coeffs}
\end{equation}
The mode $n=0$ satisfies $a_{0,0}+\sum_j a_{0,j}=0$, so that $\lambda=0$ is always an eigenvalue. It reflects the global phase-shift symmetry and is excluded from the transverse stability analysis.

For each fixed $n\neq0$, Eq.~\eqref{eq:scalar_mode} is a scalar delay differential equation with real instantaneous coefficient $a_{n,0}$ and complex delayed coefficients. Absolute stability of such equations, that is, stability for all values of the delays $\tau_j\ge0$, is characterized by Corollary~7 of Ref.~\cite{S-yanchuk2022absolute}. Since $a_{n,0}\in\mathbb{R}$, the equation is absolutely stable if and only if
\begin{equation}
a_{n,0}+\sum_{j}|a_{n,j}|\le 0,
\qquad
a_{n,0}+\sum_{j}a_{n,j}\neq 0.
\label{eq:yanchuk}
\end{equation}
Using $|a_{n,j}|=|c_j|$, the first condition becomes $-\sum_j c_j+\sum_j|c_j|\le0$, which holds if and only if
\begin{equation}
c_j\ge 0 \qquad \text{for all } j=1,\dots,N-1,
\label{eq:sign_condition}
\end{equation}
and the second condition is the nondegeneracy requirement
\begin{equation}
\sum_{j} c_j\bigl(e^{{\rm i} q_n j}-1\bigr)\neq 0
\qquad \text{for all } n=1,\dots,N-1.
\label{eq:nondeg}
\end{equation}
For attractive coupling ($w_j\ge0$, $\sigma>0$), condition \eqref{eq:sign_condition} reads $h'(j\psi_m-\Omega\tau_j)\ge0$ for all $j$, so absolute transverse stability is guaranteed when all delayed interaction channels lie on a locally attractive branch of the coupling function.

Differentiating Eq.~\eqref{eq:freq_lock} with respect to $\tau_1=1/v$ yields
\begin{equation}
\frac{{\rm d}\Omega}{{\rm d}\tau_1}
=
-\frac{\Omega \sum_{j} d_j c_j}
{1+\tau_1\sum_{j} d_j c_j},
\label{eq:dOmega}
\end{equation}
which is Eq.~(7) of the Letter. Under the sign condition \eqref{eq:sign_condition} the numerator is nonnegative and the denominator is positive, so ${\rm d}\Omega/{\rm d}\tau_1<0$ whenever some $c_j>0$. Hence absolutely stable segments of the locking branches have negative slope in the $(\tau_1,\Omega)$-plane. The converse is false, because a branch may be linearly stable for the specific delays $\tau_j=d_j/v$ without being absolutely stable, and for the ring considered here changes of stability do not in general occur at stationary points of $\Omega(\tau_1)$ \cite{S-jolly2026}. For the slow flow, Eq.~(3) of the Letter, the consequence is the one stated there. On an absolutely stable segment the slow drift $V'\propto(\Omega(V)-\nu)$ moves away from any crossing $\Omega(V)=\nu$, so such equilibria are longitudinally unstable and only the boundary states $v=v_{\min}$, $v=v_{\max}$ can be selected on these segments.

\section{Node-wise versus link-wise speeds}

The Letter takes the adaptive speeds to be node-wise, $v_{ij}=v_j$. Consider instead link-wise speeds $v_{ij}$ obeying $v_{ij}'=\varepsilon g(\theta_j',v_{ij})$ with the same presynaptic drive for all links leaving node $j$. For the sigmoidal rule, Eq.~(5), $\partial g/\partial v=-1$, so the difference of any two speeds $v_{ij}-v_{kj}$ obeys $(v_{ij}-v_{kj})'=-\varepsilon(v_{ij}-v_{kj})$ and decays as ${\rm e}^{-\varepsilon t}$, so all outgoing speeds of node $j$ contract onto a common value on the slow time scale, independently of their initial conditions, and the node-wise reduction is asymptotically exact. For the polynomial rule, Eq.~(6), the logarithmic coordinates $y_{ij}=\log(v_{ij}-v_{\min})-\log(v_{\max}-v_{ij})$ satisfy $y_{ij}'=\varepsilon(v_{\max}-v_{\min})(\theta_j'-\nu)$, which has the same right-hand side for all links leaving node $j$, so $y_{ij}-y_{kj}$ is constant in time. The speeds remain ordered but reach $v_{\min}$ or $v_{\max}$ together, and the selected boundary states are the same as for node-wise speeds. In both cases the key property is that the drive depends on the presynaptic activity only and not on the link. This fails for rules that depend on the phase relation between pre- and postsynaptic populations \cite{S-noori2020activity,S-pajevic2014role}, on postsynaptic rates \cite{S-talidou2022homeostatic}, or on activity accumulated over the link-specific transit time $d_{ij}/v_{ij}$ \cite{S-jolly2026}, and for such rules the speeds are genuinely link-specific.

\section{Time-scale separation and phase lag}

\paragraph{Dependence on $\varepsilon$.} The averaging argument of the Letter assumes $\varepsilon\ll\sigma\ll1$, and the relaxation-oscillation interpretation relies on slow drift along locking branches separated by fast transitions. We repeated the ring simulation of Fig.~3 of the Letter (blue trajectory with $\sigma=0.1$ and initial data on $\psi_0$ with $v=1/5$) for $\varepsilon$ between $10^{-5}$ and $10^{-2}$, each run lasting $40/\varepsilon$, i.e.\ several cycles. At every time we compute $Z=N^{-1}\sum_j{\rm e}^{{\rm i}(\theta_{j+1}-\theta_j)}$ (cyclic) and count the ring as close to the twisted state $\psi_m$ when $|Z|>0.95$, with $m$ given by $\arg Z=2\pi m/N$ (folded to $0\le m\le5$). Table~\ref{tab:ringeps} lists the fraction of time spent close to a locking branch after the first departure from $\psi_0$ and the number of transitions between branches (visits of at least $0.2/\varepsilon$). Up to $\varepsilon=10^{-3}$ the trajectory stays close to the branches for $75$--$92\%$ of the time and switches repeatedly among $\psi_5$, $\psi_1$, $\psi_2$ and, for $\varepsilon\le3\times10^{-4}$ within the simulated time, back to $\psi_0$. At $\varepsilon=2\times10^{-3}$ the cycle is irregular, and for $\varepsilon\ge3\times10^{-3}$ the branches are no longer tracked. At most two transitions occur, and for $\varepsilon=10^{-2}$ the network settles on $\psi_5$. The slow--fast picture of the Letter therefore requires $\varepsilon\lesssim10^{-3}=10^{-2}\sigma$.

\begin{table}
\caption{Dependence of the relaxation cycle of the ring network of Fig.~3 of the Letter on $\varepsilon$ (other parameters and initial data as for the blue trajectory, runs of length $40/\varepsilon$).}
\label{tab:ringeps}
\begin{ruledtabular}
\begin{tabular}{ccc}
$\varepsilon$ & time near a branch & transitions \\
\hline
$10^{-5}$ & $0.92$ & $44$ \\
$3\times10^{-5}$ & $0.89$ & $32$ \\
$10^{-4}$ & $0.77$ & $24$ \\
$3\times10^{-4}$ & $0.85$ & $22$ \\
$10^{-3}$ & $0.75$ & $16$ \\
$2\times10^{-3}$ & $0.36$ & $6$ \\
$3\times10^{-3}$ & $0.10$ & $1$ \\
$5\times10^{-3}$ & $0.30$ & $2$ \\
$10^{-2}$ & $0.95$ (on $\psi_5$) & $1$ \\
\end{tabular}
\end{ruledtabular}
\end{table}

\paragraph{Dependence on the phase lag.} With $h(x)=\sin(x+a)$ as in the Letter, the delayed coupling from a neighbour at distance $d_j$ enters the locking condition \eqref{eq:freq_lock} and the coefficients $c_j$ only through the argument $j\psi_m+a-\Omega d_j/v$. Increasing the lag by $\Delta a$ therefore has the same effect on this coupling term as increasing $1/v$ by $\Delta a/(\Omega d_j)$. This equivalent shift is the same for all neighbours at the same distance but differs between distances, so a change of $a$ does not shift the branches exactly. Since the weights decay as ${\rm e}^{-d_j}$, the nearest neighbours ($d_j=1$) dominate the sum, and to leading order a change of $a$ shifts the branches and their stability intervals along the $1/v$ axis by about $\Delta a/\Omega$ without changing their shape. The slow flow then produces the same boundary equilibria or relaxation cycles at shifted speeds. On the connectome, where the distances are heterogeneous, the lag does change which states are selected, as mapped in Fig.~2(b)--(e) of the Letter.

\section{Connectome data and parameter sweeps}

\begin{figure*}[!]
\centering
\includegraphics[width=\linewidth]{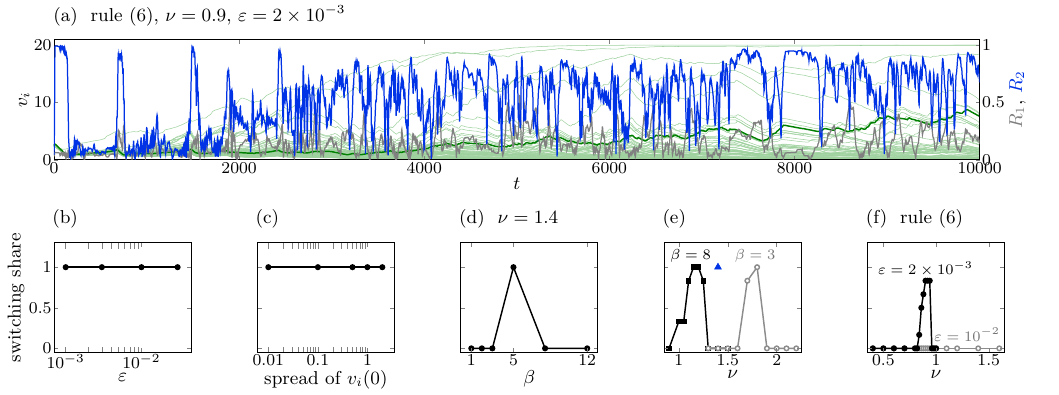}
\caption{Robustness of the switching of Fig.~1(d) of the Letter ($\sigma=1.0$, $\nu=1.4$, $h(x)=c_2[\sin(x-\pi/2)+\sin 2x]$, other parameters as in Sec.~S4). (a) Polynomial rule (6) instead of the sigmoidal rule (5), with $\nu=0.9$ and $\varepsilon=2\times10^{-3}$, showing the speeds $v_i$ (light green, with $v_1$ dark green, left axis) and order parameters $R_1$ (grey) and $R_2$ (blue, right axis). (b)--(f) Share of runs that switch, out of six random initial conditions per value, as a function of (b) the adaptation rate $\varepsilon$, (c) the standard deviation $s$ of the initial speeds $v_i(0)\sim\mathcal N(2.5,s)$, (d) the sigmoid gain $\beta$ at $\nu=1.4$, (e) the threshold $\nu$ for $\beta=8$ (black squares) and $\beta=3$ (grey circles), with $\beta=5$, $\nu=1.4$ (blue triangle), and (f) the threshold $\nu$ for rule (6) with $\varepsilon=2\times10^{-3}$ (black) and $\varepsilon=10^{-2}$ (grey). A run switches if $R_1$ or $R_2$ rises above $0.9$ and collapses below $0.1$ at least twice in the second half of the run (as in Fig.~2 of the Letter), and runs last until $t=\max(2\times10^3,20/\varepsilon)$.}
\label{fig:S1}
\end{figure*}

\paragraph{Data.} The structural connectome is obtained from Human Connectome Project diffusion data \cite{S-van2013wu} downsampled onto the 68 cortical regions of the Desikan--Killiany atlas as in Ref.~\cite{S-forrester2024whole}. It provides symmetric weights $w_{ij}$ and Euclidean inter-regional distances $d_{ij}$ (Fig.~1(a) of the Letter). In all simulations the weights are rescaled to an average row sum of $1$, so that $\sigma$ measures the total input per node, and the distances to mean $1$. The common parameters are $\omega=1$, $\beta=5$, $\varepsilon=10^{-2}$, $v_{\min}=0.1$, $v_{\max}=20$, the kernel $k=10\cdot\mathbf1_{[-0.1,0]}$, and the initial data $\theta_i\sim\mathcal U(0,2\pi)$ and $v_i\sim\mathcal N(2.5,0.1)$.

\paragraph{Sweeps of Fig.~2.} For each of the six interaction functions we integrate Eqs.~(1),(2),(5) on a $45\times45$ grid, $\sigma\in[0.3,3]$ (logarithmically spaced) and $\nu\in[1.0,1.6]$, from the same random initial condition in every cell, until $t=1.5\times10^3$. On the second half of each run we record $\langle R_1\rangle$, $\langle R_2\rangle$, the number of collapses of $R_1$ and of $R_2$ from above $0.9$ to below $0.1$, and the fraction of nodes whose mean frequency lies within $10^{-3}$ of the median. The colour coding and the switching and chimera-like criteria are those of the caption of Fig.~2. Because a single initial condition is used per cell, isolated cells near region boundaries may differ from their neighbours. The parameters of Fig.~1 were therefore chosen in the interior of the regions. For Fig.~1(b) and (d) we checked a $3\times3$ neighbourhood with four initial conditions each, and all runs at the chosen points were chimera-like and switching, respectively. For Fig.~1(c), all $24$ cells of the surrounding $5\times5$ block of the $45\times45$ grid show switching.

\paragraph{One-parameter sweeps.} Figure~\ref{fig:S1}(b)--(f) shows sweeps about the parameters of Fig.~1(d). The adaptation rate $\varepsilon$ and the spread of the initial speeds do not affect the outcome, see panels (b) and (c). The period of the switching grows as $\varepsilon^{-1}$ for $\varepsilon\le10^{-2}$. The sigmoid gain $\beta$ sets the position of the switching band in $\nu$, see panels (d) and (e), because the soft threshold keeps the speeds near $v_{\min}+(v_{\max}-v_{\min})f(\beta(\Omega-\nu))$ while $\Omega<\nu$, so a larger $\beta$ requires a smaller $\nu$ for the same balance between drift and locking ($\nu\approx1.75$, $1.4$ and $1.15$ for $\beta=3$, $5$ and $8$). The polynomial rule (6) has a hard threshold, so switching requires $\nu$ inside the range of collective frequencies attained by the cluster states ($0.55$--$1$), in fact close to its upper end, see panel (f). There, and for slow enough adaptation ($\varepsilon=2\times10^{-3}$), it produces the same recurrent switching between cluster states as the sigmoidal rule, as panel (a) shows, so the mechanism does not depend on the specific form of the adaptation rule. For $\varepsilon=10^{-2}$ the cluster states break up only partially and no run meets the switching criterion. With rule (6) the speeds spread out towards both bounds, in line with Sec.~S2.

\paragraph{Persistence of the chimera state.} The simulation of Fig.~1(b) ($\sigma=0.6$, $\nu=1.44$, $h=\sin$) was continued to $t=10^6$ with unchanged parameters. In each of the $100$ consecutive windows of $10^4$ time units the time-averaged order parameter stays between $0.84$ and $0.86$ and the dispersion of the node mean frequencies between $8.1\times10^{-2}$ and $9.0\times10^{-2}$, i.e.\ the network remains partially coherent with a group of oscillators drifting at their own frequencies. The number of nodes locked to the common frequency (within $10^{-3}$) varies between $33$ and $45$ of $68$ from window to window, so the coherent group slowly exchanges members while the chimera character of the state persists. The mean speed fluctuates between $0.66$ and $0.72$. We found no sign of collapse to synchrony or to complete incoherence over $10^4$ adaptation times.

\end{document}